\documentclass[%
twocolumn,
trackchanges,
tighten,
amsmath,
astrosymb,
]{aastex7}

\def\alfven{Alfv\'en}
\def\ampere{Amp\`ere}
\def\b{\mathbf}
\def\p{\partial}
\def\d{\mathrm{d}}
\def\los{\mathrm{LOS}}
\def\r{\mathrm{RM}}
\def\dm{\mathrm{DM}}

\renewcommand{\Re}{\operatorname{Re}}

\begin{document}

\title{Rotation Measure Substructures Induced by the Ponderomotive Force of Inertial \alfven~Waves}

\author[orcid=0009-0007-2536-5080,sname='Zhao']{Qing Zhao}
\affiliation{
    Purple Mountain Observatory, Chinese Academy of Sciences, Nanjing 210023, China; dxiao@pmo.ac.cn
}
\affiliation{
    School of Astronomy and Space Sciences, University of Science and Technology of China, Hefei 230026, China
}
\email[hide]{qingzhao@pmo.ac.cn}

\author[orcid=0000-0002-4304-2759,sname='Xiao']{Di Xiao*}
\affiliation{
    Purple Mountain Observatory, Chinese Academy of Sciences, Nanjing 210023, China; dxiao@pmo.ac.cn
}
\affiliation{
    State Key Laboratory of Radio Astronomy and Technology, Purple Mountain Observatory, Chinese Academy of Sciences, 10 Yuanhua Road, Nanjing 210023, China
}
\email{dxiao@pmo.ac.cn}

\author[orcid=0000-0002-6299-1263,sname='Wu']{Xue-Feng Wu}
\affiliation{
    Purple Mountain Observatory, Chinese Academy of Sciences, Nanjing 210023, China; dxiao@pmo.ac.cn
}
\affiliation{
    School of Astronomy and Space Sciences, University of Science and Technology of China, Hefei 230026, China
}
\email[hide]{xfwu@pmo.ac.cn}

\begin{abstract}

The rotation measure (RM) and dispersion measure (DM) of fast radio bursts (FRBs) serve as critical probes of the magneto-ionic environments along the line of sight. The significant temporal evolution of RM observed in some repeating FRBs is generally attributed to the local environment of the source, since the intergalactic medium is not expected to vary on such short timescales. Recent observations of repeating FRB 20201124A and FRB 20220529 exhibit complex RM phenomenology, including large-amplitude global fluctuations and short-term substructures. Here, we attribute these short-term RM variations to the ponderomotive force exerted by inertial \alfven~waves (IAWs). We propose that IAWs, generated via magnetic reconnection or turbulent cascades in a low-$\beta$ plasma, induce nonlinear density perturbations in the source environment. We demonstrate that the resulting plasma density redistribution can produce RM suppression consistent with observed substructures. 
This model presents a physically motivated mechanism for the short-term RM variability observed in active repeaters. It demonstrates that such fluctuations can arise from wave-driven density cavitation within a broad, coupled parameter space involving wave amplitude, plasma density, and temperature, thereby characterizing the localized plasma dynamics required to produce the observed RM jitters.

\end{abstract}

\section{Introduction} 
Fast radio bursts (FRBs) represent a class of millisecond-duration radio transients characterized by extremely high brightness temperatures. Since the discovery of the archetypal ``Lorimer burst" \citep{2007Sci...318..777L}, the high dispersion measures (DMs) of these events—far exceeding the Galactic contribution—have firmly established their cosmological origin. Over the past decade, the rapidly growing sample of FRBs, particularly the continuous monitoring of repeating sources, has provided critical insights into their radiation mechanisms and origins \citep[for recent reviews, see e.g., ][]{2021SCPMA..6449501X,2022A&ARv..30....2P,2023RvMP...95c5005Z}.

While the extragalactic distances of FRBs challenge direct imaging of their sources, propagation effects offer a unique probe of the local environment. The discovery of periodic activity in certain repeaters \citep{2020Natur.582..351C, 2020MNRAS.495.3551R,2021MNRAS.500..448C,2025arXiv250715790W} suggests that these sources may reside in binary systems \citep{2020ApJ...893L..26I,2020ApJ...893L..39L,2020MNRAS.498L...1Z}. In this framework, the significant secular rotation measure (RM) evolution observed in active sources, such as FRB 20201124A \citep{2022Natur.609..685X} and FRB 20190520B \citep{2023Sci...380..599A}, is naturally interpreted as propagation through the accretion disk or the dynamic wind of a massive companion \citep{2022NatCo..13.4382W}. More recently, detailed monitoring of FRB 20201124A and FRB 20220529 \citep{doi:10.1126/science.adq3225} has revealed not only significant RM evolution but also a potential periodicity in these magnetic variations \citep{2025arXiv250506006X, 2025arXiv250510463L,zhang2025periodicactivityepochsfrb}. Such periodic magnetic variability points to a structured, recurring environment, likely driven by stellar activity or orbital motion surrounding the source.

While the binary scenario successfully reproduces the secular RM evolution, high-cadence observations reveal complex short-term variability superimposed on this smooth orbital trend. These fluctuations are frequently attributed to random turbulence in the local medium, yet such explanations often remain phenomenological and lack a quantitative physical description. A more specific scenario involves magnetar flares launching ejecta; as the FRB propagates through the expanding shell, the RM is predicted to decrease and recover over timescales of days to weeks \citep{2025A&A...698L...3X}. Indeed, such distinct features have been observationally confirmed: FRB 20201124A exhibited sharp RM drops ranging from $-30$ to $-100 \, \text{rad m}^{-2}$, followed by a recovery within just 2--3 days. Analogous rapid ``dips" were also identified in FRB 20220529. These short-term substructures imply the transient formation of plasma density depletions along the line of sight. In this work, we investigate a mechanism to generate such structures via wave-particle interactions. We propose that such density perturbations are induced by the nonlinear ponderomotive force exerted by small-scale inertial \alfven~waves (IAWs) in the cold plasma regime.

Large-amplitude \alfven{ic} fluctuations are a generic feature of magnetized, collisionless outflows, extending well beyond the solar wind context \citep{1971JGR....76.3534B,1995SSRv...73....1T,2013LRSP...10....2B}. In the environments of massive stars and neutron star binaries, similar but more extreme conditions—such as high flow velocities, strong shear, and pronounced magnetic inhomogeneities—naturally arise \citep{1969ApJ...157..869G,1993ApJ...403..249A,2009ASSL..357..421K}. These conditions are particularly prevalent in interaction regions where stellar winds encounter magnetospheres or opposing winds, leading to compression layers and shear-driven instabilities \citep{2008MNRAS.387...63B}. Consequently, broadband, nonsinusoidal \alfven{ic} fluctuations are generated and amplified, analogous to those observed in heliospheric plasmas \citep{1971JGR....76.3534B,1989JGR....9411739T,2018LRSP...15....1R}. 

As these fluctuations cascade toward smaller transverse scales, they develop strong anisotropy, favoring filamentary, field-aligned structures \citep{2000SSRv...92..423S,2009ApJS..182..310S}. At kinetic scales in low-$\beta$ plasmas where the \alfven~speed exceeds the electron thermal speed ($v_{\rm A} \gtrsim v_{Te}$), electron inertia becomes dynamically significant. It mediates the transition from large-scale \alfven~waves to inertial \alfven~waves (IAWs), which are characterized by finite parallel electric fields \citep{1984P&SS...32.1387G,1996JGR...101.5085L,2000SSRv...92..423S}. These IAWs can be excited via linear processes, such as magnetic field-aligned electron beams \citep{1999LNP...536....1S} or sheared magnetic fields, or through nonlinear processes driven by large-amplitude, high-frequency electromagnetic or electrostatic drivers \citep{1998SoPh..182..411V}. 

\begin{figure} [ht]
    \centering
        \includegraphics[width=0.48\textwidth,height=0.25\textwidth]{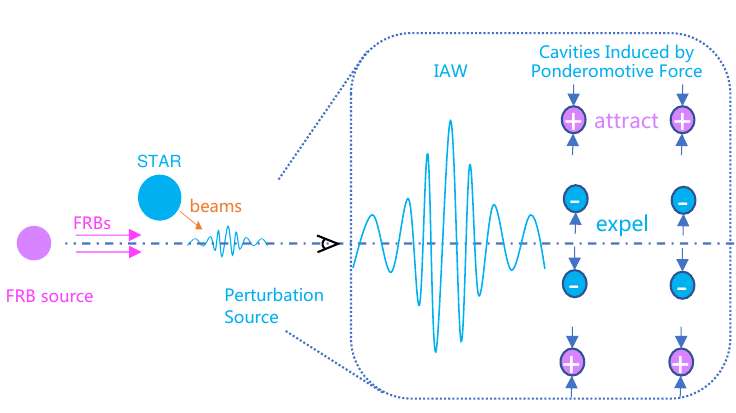}
    \caption{
       Schematic illustration of the physical mechanism driving short-term RM substructures. The FRB source is embedded in a binary system with a massive companion. Inertial \alfven~Waves (IAWs) are excited within the local environment, either by electron beams launched from the companion or through the turbulent cascade of large-scale \alfven~wave initiated by magnetic reconnection in the stellar wind. The FRB emission and IAWs are assumed to propagate parallel to the line of sight (LOS), indicated by the dashed line. The resulting nonlinear ponderomotive force acts to expel electrons from the wave packets, forming density cavities that manifest as rapid fluctuations (dips) in the observed RM.
    }
    \label{fig:schematic}
\end{figure}

This letter is organized as follows. In Section~\ref{sec:methods}, we derive the theoretical formalism for RM variations induced by the ponderomotive force of IAWs. In Section~\ref{sec:results}, we apply this model to interpret the RM substructures observed in active repeaters. Finally, a discussion and summary are presented in Section~\ref{sec:discussion}.

\section{Methods} \label{sec:methods}

IAWs can be excited by electron beams via instabilities or arise from the cascade of large-scale \alfven~wave initiated by magnetic reconnection (e.g., in the magnetotail regions). We assume a geometry where the line of sight (LOS) is parallel to the background magnetic field $\mathbf{B}_0$, along which the IAWs propagate (see Figure \ref{fig:schematic}). In low–$\beta$ collisionless plasmas, both laboratory experiments and space observations confirm that collisionless (Landau) damping of IAWs is negligible, consistent with the inertial-limit dispersion relation \citep{2010PhRvL.104i5001K}.
Adopting characteristic parameters for the source environment—electron temperature $T_e=10^3~\text{K}$, ambient electron density $n_{e0} = 10^4 \, \text{cm}^{-3}$, and magnetic field $B_0 = 10^{-2} \, \text{G}$—the corresponding electron thermal speed and \alfven~speed are $v_{Te}=1.7\times10^7~\text{cm}~\text{s}^{-1}$, $v_\text{A} \approx 2.2 \times 10^7 \, \text{cm s}^{-1}$. The wave frequency is estimated as $\omega = k_\parallel v_\text{A}/(1+k_\perp^2\lambda_e^2)^{1/2} \sim 1 \, \text{rad s}^{-1}$ where $k_\perp\lambda_e\sim1$ denotes the characteristic transverse scale of IAW (see Appendix \ref{appendix:dispersion} for the detailed dispersion relation) .

Linear Vlasov theory in the inertial regime demonstrates that the damping rate $\gamma$ remains much smaller than the real frequency $\omega$ ($\gamma \ll \omega$) over a wide parameter range. Typically, the damping rate satisfies $\gamma\propto\omega \exp\left({-v_\text{phase}^2/2v_{Te}^2}\right)$, where the phase velocity $v_\text{phase}=v_\text{A}/\sqrt{1+k_\perp^2\lambda_e^2}$, implying that IAWs experience only weak collisionless damping \citep{1996JGR...101.5085L,1998SoPh..182..411V}. Consequently, the characteristic Landau damping timescale $\tau_{\text{LD}} \approx \gamma^{-1}$ is of the order of $10^3-10^4$ s (several hours). This timescale far exceeds the FRB duration and propagation time through the region, implying that IAW structures can persist stably during the interaction.

The observed RM of an extragalactic FRB comprises contributions from distinct plasma components along the LOS \citep[e.g.,][]{2021SCPMA..6449501X,2023RvMP...95c5005Z,2023MNRAS.520.2039Y}, i.e., 
\begin{eqnarray}
    \r_\text{obs} = &\r&_\text{ion} + \r_\text{MW} + \r_\text{IGM} + \frac{\r_\text{host}}{(1+z_\text{host})^2} \nonumber\\
    &+& \frac{\r_\text{loc}}{(1+z_\text{host})^2},
    \label{rm:obs}
\end{eqnarray}
where $z_\text{host}$ denotes the redshift of the host galaxy. The terms $\r_\text{ion}$, $\r_\text{MW}$, $\r_\text{IGM}$, $\r_\text{host}$ and $\r_\text{loc}$ represent contributions from the Earth's ionosphere, the Galactic interstellar medium (ISM), the intergalactic medium, the host galaxy's ISM, and the local environment surrounding the FRB source, respectively. Given that the Galactic, intergalactic, and large-scale host ISM components are unlikely to exhibit variations on short timescales, the rapid RM evolution observed in repeaters is attributed to the dynamic local environment. For the remainder of this work, we focus on this variable component and refer to $\r_{\text{loc}}$ simply as RM.

For a cold, non-relativistic magneto-ionic medium, the RM is defined by the integral of the electron density $n_e$ and the LOS magnetic field $B_\parallel$. Numerically, this is given by:
\begin{equation}
    \r \approx 8.1 \times 10^5 \int_{\text{LOS}} \left( \frac{n_e}{\text{cm}^{-3}} \right) \left( \frac{B_\parallel}{\text{G}} \right) d\left(\frac{l}{\text{pc}}\right) \, \text{rad m}^{-2}.
    \label{eq:RM_def}
\end{equation}

{
We next consider the low-$\beta$ regime ($\beta < m_e/m_p$), where the \alfven~speed exceeds the thermal speeds of electrons and ions ($v_\text{A} > v_{Te}, v_{Tp}$). In this regime, field-aligned electron dynamics are inertia-dominated rather than pressure-driven, as the parallel electric force far exceeds the thermal gradient ($\nabla P_j = \nabla\left(n_j T_j\right)$). Since the parallel acceleration scales as $\partial_t u_\parallel = q_j E_\parallel/m_j \propto 1/m_j$, the field-aligned current is carried predominantly by electrons (see Appendix \ref{appendix:dispersion}). The large amplitude of IAWs, combined with transverse scales comparable to the electron skin depth ($k_\perp \lambda_e \sim 1$), naturally generates a significant ponderomotive force on species $j$ given by \citep{1996JGR...101.5085L,2000SSRv...92..423S}
\begin{eqnarray}
    &&\b{F}_j = \frac{1}{2}q_j\nabla \left\langle \b{r}_1 \cdot \b{E}_1 \right\rangle,
\end{eqnarray}
where the displacement vector $\mathbf{r}_1$ is
\begin{eqnarray}
    &&\b{r}_1 = -\frac{q_j}{m_j}\left[\frac{\omega_{cj}\b{\hat{z}} \times \dot{\b{E}}_1}{\omega^2-\omega_{cj}^2} + \frac{\b{\hat{z}} \times \left(\b{E}_1 \times \b{\hat{z}}\right)}{\omega^2-\omega_{cj}^2} + \frac{\left(\b{\hat{z}} \cdot \b{E}_1\right)\b{\hat{z}}}{\omega^2}\right]. \nonumber\\
\end{eqnarray}
Assuming a harmonic field $\b{E}_1 = \Re\left[\b{E}(\b{x})\exp(-i\omega t)\right]$, we obtain
\begin{eqnarray}
    \b{F}_j = -\frac{q_j^2}{4m_j}\nabla\left[\frac{\omega_{cj}\b{\hat{z}} \cdot \left(\dot{\b{E}}_\perp \times \b{E}_\perp^*\right)}{\omega^2-\omega_{cj}^2} + \frac{\left\vert \b E_\perp\right\vert^2}{\omega^2-\omega_{cj}^2} + \frac{\left\vert E_z\right\vert^2}{\omega^2}\right]. \nonumber\\
    \label{pond}
\end{eqnarray}
As demonstrated by \citet{1998PhRvL..80.3523B}, the electron ponderomotive force is $\b{F}_e \simeq -(n_e q_e^2/4m_e)\nabla (\left\vert E_z\right\vert^2/\omega^2)$. For a cylindrical mode structure, this force acts to expel electrons from the axis ($r=0$) where the parallel field intensity (scaling as $J_0^2(k_\perp r)$) is maximized. Conversely, the ion ponderomotive force is $\b{F}_i \simeq -(n_i q_i^2/4m_i)\nabla (\left\vert E_r\right\vert^2/(\omega^2-\omega_{ci}^2)) \simeq (n_i q_i^2/4m_i)\nabla (\left\vert E_r\right\vert^2/(\omega_{ci}^2))$, which tends to attract ions to the node at $k_\perp r \simeq 2$ where $J_1^2(k_\perp r)$ is at a maximum. However, comparing the magnitudes reveals that the ion force is approximately an order of magnitude weaker than that of the electrons. Consequently, the net effect is the formation of a significant electron density cavity at the wave center, bounded by a minor density ridge. 

\begin{figure} [ht]
    \centering
        \includegraphics[width=0.48\textwidth,height=0.46\textwidth]{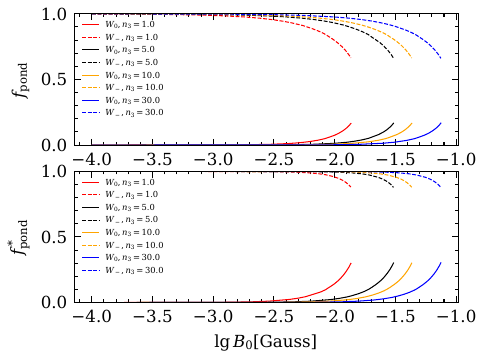}
    \caption{
        Dependence of the ponderomotive filling factors $f_{\text{pond}}$,  $f^*_{\text{pond}}$ on the ambient magnetic field strength for various electron densities ($n_{e0} = \{1, 5, 10, 30\} \times 10^3 \, \text{cm}^{-3}$). In both panels, the solid and dashed curves correspond to the principal ($W_0$) and secondary ($W_{-1}$) branches of the Lambert function solution, respectively. All calculations assume a fixed perturbation amplitude of $\alpha = 10^{-4}$.
    }
    \label{fig:fpond}
\end{figure}

Assuming a quasi-static equilibrium, the fluid equations of motion along the magnetic field reduce to a balance between the ponderomotive force, the ambipolar electric field, and the thermal pressure gradient. The parallel component equations for electrons and ions are given by:
\begin{eqnarray}
    -\frac{n_e q_e^2}{4m_e}\frac{\p}{\p s} \frac{\left\vert E_z\right\vert^2}{\omega^2} - n_e q_e \frac{\p \phi}{\p s} - \frac{\p}{\p s}\left(n_e k_B T_e\right)=0,\\
    \frac{n_i q_i^2}{4m_i}\frac{\p}{\p s} \frac{\left\vert E_r\right\vert^2}{\omega_{ci}^2} - n_i q_i \frac{\p \phi}{\p s} - \frac{\p}{\p s}\left(n_i k_B T_i\right)=0,
\end{eqnarray}
where $s$ denotes the coordinate along the magnetic field line, and $\phi$ represents the self-consistent ambipolar potential to be determined. Integrating these equations yields the distributions for the species densities:
\begin{eqnarray}
    n_e = n_{e0}\exp\left[-\left(e^2\left\vert E_z\right\vert^2/4m_e\omega^2 - e\phi\right)/k_B T_e\right], \nonumber\\
    n_i = n_{i0}\exp\left[\left(e^2\left\vert E_r\right\vert^2/4m_i\omega_{ci}^2 - e\phi\right)/k_B T_i\right],
    \label{ne:ni}
\end{eqnarray}
with $q_e=-e$ and $q_i=+e$ being substituted. Imposing the quasi-neutrality condition ($n_e \approx n_i$) allows us to solve for the self-consistent ambipolar potential $\phi$: 
\begin{eqnarray}
    \phi = \frac{e}{4m_e\left(T_e + T_i\right)}\left(\frac{\left\vert E_z\right\vert^2T_i}{\omega^2} + \frac{m_e\left\vert E_r\right\vert^2T_e}{m_i\omega_{ci}^2}\right),
    \label{ambi}
\end{eqnarray}
Combining Eq.~(\ref{ambi}) with Eq.~(\ref{ne:ni}), we obtain the electron density
\begin{eqnarray}
    &&n_e = n_{e0}\exp\left[-\frac{e^2}{4m_e k_B (T_e + T_i)} \left(\frac{\left\vert E_z\right\vert^2}{\omega^2} - \frac{m_e\left\vert E_r\right\vert^2}{m_i\omega_{ci}^2}\right)\right]. \nonumber\\
\end{eqnarray}

As derived in Appendix A (Eq.~(\ref{current:para})), using $\omega_{pe}^2 = n_e q_e^2/\epsilon_0 m_e$, we see that $\left\vert E_z/\omega\right\vert = c^2 \mu_0 \left\vert J_z\right\vert/\omega_{pe}^2$. Focusing on the wave axis ($r \to 0$) where the perpendicular field $E_r$ vanishes, the ambipolar potential simplifies to: 
\begin{eqnarray}
    \phi \approx \frac{m_e \left\vert J_z\right\vert^2}{4 n_e^2 e^3 \left(1+T_e/T_i\right)}.
\end{eqnarray}
Consequently, the electron density near the core is governed by:
\begin{eqnarray}
    \frac{n_e}{n_{e0}} = \exp\left(-\frac{e\phi}{k_B T_i}\right) = \exp\left[-\frac{m_e \left\vert J_z\right\vert^2}{4n_e^2 e^2 k_B \left(T_e + T_i\right)}\right]. \nonumber\\
    \label{ne:cav}
\end{eqnarray}
Defining a dimensionless nonlinearity parameter $\xi = m_e\left\vert J_z\right\vert^2/4n_{e0}^2 e^2 k_B \left(T_e + T_i\right)$, we arrive at the final implicit relation for the density perturbation:
\begin{equation}
    \frac{n_e}{n_{e0}} = \exp \left( -\xi \frac{n_{e0}^2}{n_e^2} \right).
    \label{eq:transcendental}
\end{equation}
This result aligns with the analysis of \citet{1998PhRvL..80.3523B}. {To understand the behavior of the solution, it is instructive to invert Eq. (\ref{eq:transcendental}) as $\xi = (n_e/n_{e0})^2 \ln(n_{e0}/n_e)$. Analysis shows that for a given $\xi < 1/(2e) \approx 0.18$, there exist two possible solutions for $n_e/n_{e0}$. One solution corresponds to a small perturbation ($n_e \approx n_{e0}$), while the other corresponds to a deep density cavity ($n_e \ll n_{e0}$). Physically, we focus on the branch connected to the unperturbed state ($n_e \to n_{e0}$ as $\xi \to 0$), which is described by the principal branch of the Lambert W function ($W_0$). On this branch, as the current intensity ($\xi$) increases, the density $n_e$ decreases monotonically. Crucially, the $n_e^{-2}$ dependence in the exponent creates a strong nonlinear feedback loop: a reduction in density increases the effective potential barrier, which further expels electrons. This self-amplifying mechanism allows for the formation of deep density cavities even with moderate field-aligned currents, directly accounting for the rapid ``dip" substructures observed in the RM evolution.}

To quantify the density depletion, we estimate the parallel current density $J_z$ via \ampere's law, assuming a specific magnetic field perturbation structure. We consider a single IAW packet characterized by a plane wave modulated by a Gaussian envelope. The azimuthal magnetic perturbation $\delta B_{\theta}$ is modeled as:
\begin{eqnarray}
    \delta B_{\theta}(\omega) =& \alpha & B_{\theta 0} \sin(k_\parallel z-\omega t +\phi)\sin(k_\perp r)
    \nonumber\\
    &\cdot&
    \exp\left[-\left(r^2/\lambda_e^2+z^2/L^2\right)\right],
    \label{mag:perb}
\end{eqnarray}
where $\alpha$ denotes the perturbation amplitude relative to the ambient field $B_0$, and $\lambda_{e,\parallel}$, $L$ denote the characteristic transverse and parallel scales of the Gaussian wavefront respectively.
 
To address the physical origin of the IAWs and their associated magnetic perturbations in our model, we draw upon high-resolution {\textit{in-situ}} measurements from space plasmas. Observations from the \textit{Freja} satellite---specifically designed to resolve micro-physics within small-scale \alfven{ic} features in the topside ionosphere \citep{1994SSRv...70..405L}---provide unequivocal evidence for widespread \alfven~wave turbulence. In these strongly magnetized, low-$\beta$ environments, electromagnetic fluctuations in the $\sim$ 1-7 Hz frequency range exhibit an electric-to-magnetic field perturbation ratio of $\delta{E}/\delta{B}\simeq{v_{\text{A}}}$, characteristic of IAWs \citep{1998JGR...103.4315S}. Crucially, the observational data reveal a strong spatial correlation between these transverse magnetic irregularities---interpreted as inertial \alfven~resonance cones---and sharp, large-scale electron density cavities \citep{1998PhRvL..80.3523B}. This demonstrates that the IAW-driven ponderomotive force is a fundamentally robust mechanism capable of heating and evacuating plasma along magnetic field lines, providing an existence proof for the density depletion process modeled in this work.

Rather than treating the relative wave amplitude $\alpha$ as an arbitrary free parameter or a universal constant, its magnitude is framed as a \textbf{parameterized fiducial value}. We acknowledge that applying measurements from the Earth's auroral zone to astrophysical environments involves inherent assumptions about the turbulent cascade---specifically, how relative wave intensities extrapolate across the vast separation between macroscopic injection scales and microscopic kinetic scales. Since the specific driving mechanisms (e.g., reconnection vs. shear flows) and the extent of scale separation in FRB environments remain observationally unconstrained, we do not treat $\alpha$ as a strictly fixed quantity. Instead, we adopt $\alpha = 10^{-4}$ as a representative baseline inspired by the only available \textit{in-situ} analogs \citep{2001PCEC...26..201C}. To ensure the robustness of our conclusions against these environmental uncertainties, we perform a comprehensive sensitivity analysis across the $(\log B_0, \log \alpha)$ parameter space (see Figure \ref{fig:fpond-contour}), which confirms that IAW-driven density cavitation remains a robust outcome over a broad, coupled regime of plasma conditions.

\begin{figure} [ht]
    \centering
        \includegraphics[width=0.48\textwidth,height=0.46\textwidth]{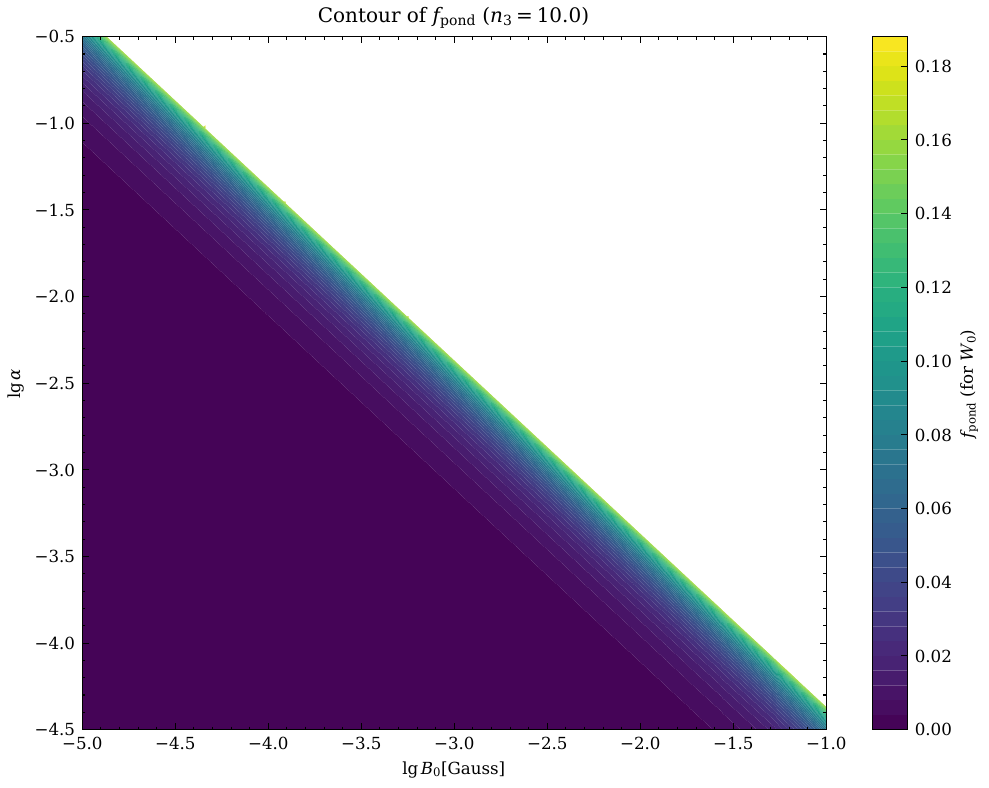}
    \caption{
        Contour plot of the ponderomotive filling factor $f_{\text{pond}}$ (evaluated for the principal $W_0$ branch) as a function of the ambient magnetic field $B_0$ and the perturbation amplitude $\alpha$. The background electron density is fixed at a fiducial value of $n_{e0} = 10^4 \, \text{cm}^{-3}$. The color scale indicates the magnitude of $f_{\text{pond}}$. The unshaded regions denote parameter regimes excluded by the physical constraint  $\alpha B_1/T_4^{1/2}n_3^{1/2} \leq 1.1\times10^{-5} $, which ensures the existence of real-valued plasma density solutions.
    }
    \label{fig:fpond-contour}
\end{figure}

The transcendental Eq. (\ref{eq:transcendental}) can be solved analytically using the Lambert W function, $W(z)$, defined by $z = W(z)e^{W(z)}$. The resulting magnitude of electron density depletion is: 
\begin{eqnarray}
    \Delta n_e = n_{e0} - n_{e} = n_{e0}\left[1-e^{W\left(-2\xi\right)/2}\right] . \nonumber\\
\end{eqnarray}
Note that for physically meaningful (real-valued) density solutions, the nonlinearity parameter must satisfy the condition $0 < \xi \le 1/(2e)$. The observed RM can be decomposed as $\rm RM = RM_{\text{base}} + \Delta RM$, where $\rm RM_{\text{base}}$ represents the baseline contribution from the unperturbed background ($n_{e0}$). The contribution from the density cavity is then:
\begin{eqnarray}
    \Delta \r = -8.1\times10^{5}\cdot f_\mathrm{pond}\left(\xi\right)\cdot\left\langle n_{e0} B_{\parallel}\right\rangle L\quad\mathrm{rad}\cdot\mathrm{m}^{-2},\nonumber\\
\end{eqnarray}
where $f_{\text{pond}}(\xi)$ is the dimensionless filling factor:
\begin{eqnarray}
    f_\mathrm{pond}&\left(\xi\right)=
     \frac{1}{L}\int_\mathrm{SS} \left[1-e^{W\left(-2\xi\right)/2}\right] \d l.
    \label{factor} 
\end{eqnarray}
Here, the subscript SS denotes the integration path across a Single Structure (specifically, an individual IAW packet) within the local plasma environment. $L$ represents the characteristic path length of this single structure. This definition isolates the contribution of an individual density cavity, acknowledging that the total observed $\Delta \rm RM$ may result from the cumulative effect of multiple such structures along the line of sight.  Evaluating the nonlinearity parameter $\xi$ with normalized units, we obtain:
\begin{eqnarray}
    \xi = T_4^{-1}n_3^{-2}\cdot \left(\frac{\left|\frac{1}{r}\frac{\p}{\p r}\left(r\delta{B_{\theta}}\right)\right|}{1.6\times10^{-8}~\mathrm{G}\cdot\mathrm{cm}^{-1}}\right)^2,
    \label{xi}
\end{eqnarray}
where $T_4$ is measured in units of {$10^4~\mathrm{K}$}, $n_3$ in units of $10^3~\mathrm{cm}^{-3}$.

\begin{figure*} [ht]
    \centering
        \includegraphics[width=1.0\textwidth,height=0.66\textwidth]{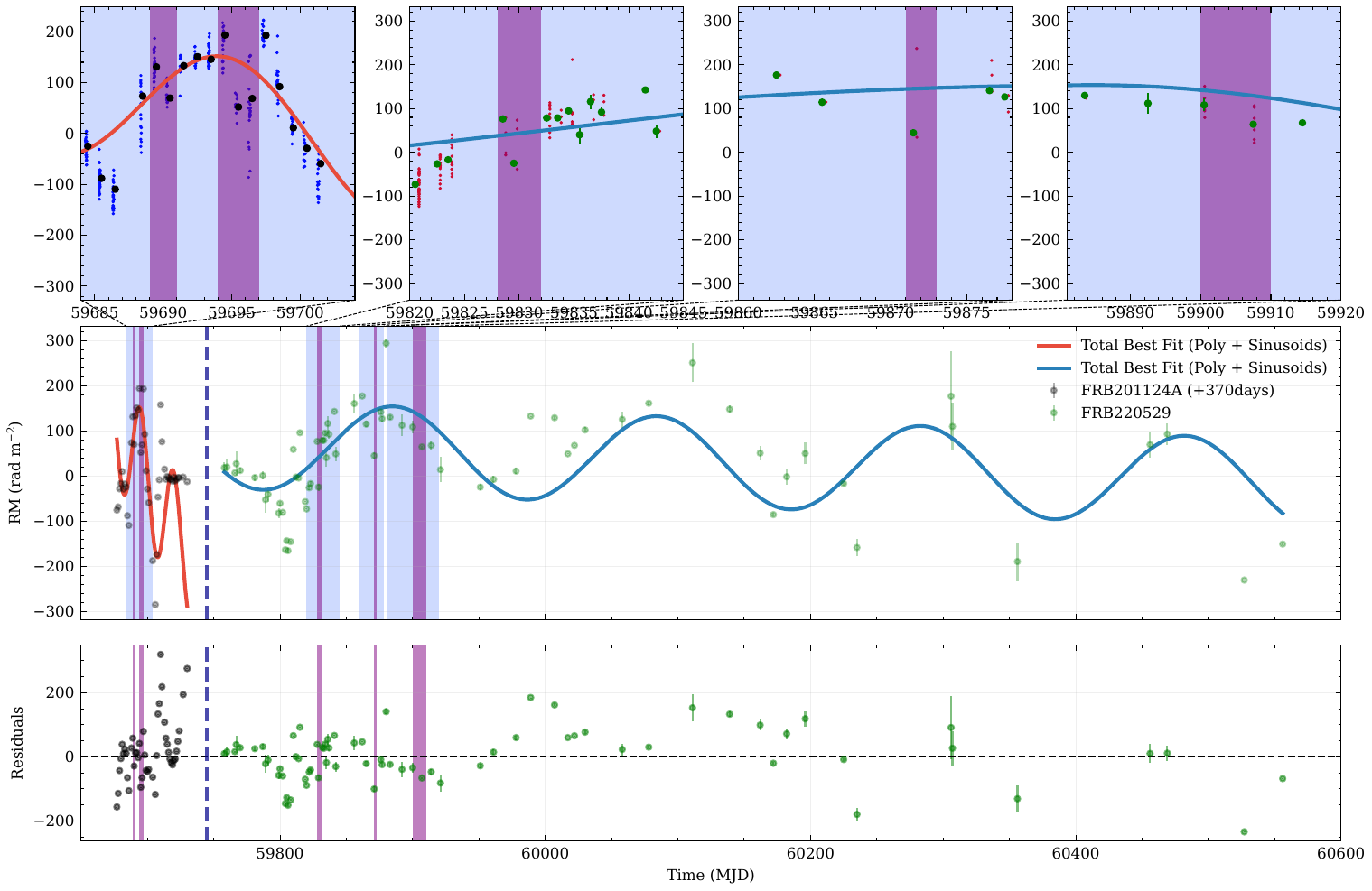}
    \caption{
        The RM evolution in FRB 20201124A (+370 days) and FRB 20220529, separated by the vertical dashed line. The blue and red points represent the observed RM values, while the black and green points denote the daily-averaged RM values. The red and blue curves show the fit results using a composite model consisting of a first-order polynomial and a sinusoidal component. The corresponding residuals are displayed in the lower panels. The vertical purple shaded bands indicate the selected RM substructures where multiple day-scale dips are observed.
        Temporal RM evolution of FRB 20201124A (+370 days) and FRB 20220529, separated by the vertical dashed line. The scatter points (blue and red) represent the individual observed RM values, while the solid points (black and green) denote the daily-averaged measurements. The solid curves (red and blue) indicate the best-fit secular trends, modeled as a composite of a first-order polynomial and a sinusoidal component. The corresponding residuals are displayed in the lower panels. The vertical purple shaded bands highlight epochs of rapid RM suppression, ranging from isolated single-dip events (narrow bands) to extended intervals hosting multiple day-scale fluctuations (wide band).
    }
    \label{fig:rm201124}
\end{figure*}

Similarly, we consider the impact on the DM and free-free optical depth. The $\dm$ contributed by a cold plasma block is
\begin{eqnarray}
    \dm = \int_\los n_e \d l,
\end{eqnarray}
and the corresponding optical depth
\begin{eqnarray}
    \tau = 8.2\times10^{-2}T^{-1.35}\nu^{-2.1}\int_\mathrm{LOS} n_e^2 \d l
\end{eqnarray}
should be less than unity. The DM correction induced by the cavity is $\Delta\dm = -f_\mathrm{pond}(\xi) \cdot \dm_{\mathrm{base}}$. Observationally, the DM of FRB 20201124A and FRB 20220529 remains remarkably stable \citep{2022Natur.609..685X,doi:10.1126/science.adq3225}. This stability implies that the baseline DM contribution from the local environment ($\dm_{\mathrm{base}}$) is likely small compared to the total extragalactic DM. Consequently, even a significant fractional depletion $f_{\text{pond}}$ results in a $\Delta\rm DM$ that is negligible relative to the total observed value and its measurement uncertainty. In contrast, the optical depth correction involves a different weighting factor due to the quadratic density dependence ($n_e^2$):
\begin{eqnarray}
    \Delta\tau = -f^*_\mathrm{pond}(\xi) \cdot \tau_{\mathrm{base}},
\end{eqnarray}
where the quadratic filling factor is given by: $f^*_\mathrm{pond}(\xi) = \frac{1}{L}\int_\mathrm{SS} \left[1-e^{W\left(-2\xi\right)}\right] \d l$.

\section{Results} \label{sec:results}

We consider a binary system comprising an FRB-emitting magnetar and a massive stellar companion exhibiting strong outflows. The radial profiles of the electron temperature $T_e$, electron density $n_e$, and ambient magnetic field $B$ are modeled as power-law functions of the distance $r$ from the massive star:
\begin{eqnarray}
    T_e\propto r^{-\beta_1},~n_{e}\propto r^{-\beta_2},~B\propto r^{-n},
    \label{scaling}
\end{eqnarray}
with indices $\beta_1=2/3,~\beta_2=2,~n={1}$ (consistent with a standard stellar wind model). Consequently, the parameter $\xi$ scales as $r^{\beta_1 + \beta_2 - 2n}$. Under the low-$\beta$ condition,
\begin{eqnarray}
    \beta \equiv 8\pi n_e k_B T_e / B^2 \ll m_e / m_p,
    \label{lowbeta}
\end{eqnarray}
and using the scaling relations above, the local plasma beta evolves as $\beta\left(r\right) \propto r^{-2/3}$. We therefore define a critical radius for the IAW regime as
\begin{eqnarray}
    r_\text{IAW} = r_0 \left(\frac{\beta_0}{m_e/m_p}\right)^{3/2},
    \label{rIAW}
\end{eqnarray}
where $\beta_0 \equiv \beta \left(r_0\right)$ is the plasma beta at a reference radius $r_0$. For $r>r_\text{IAW}$, the plasma naturally satisfies the conditions for IAW generation and propagation.

Next, we evaluate the factor $f_\mathrm{pond}$ and the resulting RM correction using parameters representative of the stellar outflow. Without loss of generality, we assume that the direction of ambient magnetic field is aligned with the $z$-axis. At $r=0$, the $z$-component of $ |\nabla \times \delta \mathbf{B}|^2$ is given by
 \begin{equation}
    \left|\frac{1}{r}\frac{\p}{\p r}\left(r\delta {B_{\theta}}\right)\right|^2_{r=0} = \sum_\omega 4\alpha^2k_\perp^2B_{\theta0}^2\sin^2(k_\parallel z)e^{-2z^2/L^2}.
    \label{curlBsquare}
\end{equation}
The characteristic length scales—the electron inertial length $\lambda_e$ and the parallel wavelength $\lambda_\parallel$—are 
$\lambda_e = \frac{c}{\omega_{pe}} = 2\times10^{4} n_3^{-1/2}~\mathrm{cm}$, 
$ \lambda_\parallel = \frac{2\pi}{k_\parallel} = 4.4\times10^{10}\omega_1^{-1} B_1 n_{3}^{-1/2}~\mathrm{cm}$ respectively,
where $\omega_1$ is the frequency in Hz and $B_1$ is the magnetic field in Gauss. The validity of the Lambert W solution requires the nonlinearity parameter to satisfy $\xi \le (2e)^{-1}$. Combining this condition with Eqs. (18) and (25), we derive an upper bound on the perturbation parameters as $\alpha B_1/T_4^{1/2}n_3^{1/2} \leq 1.1\times10^{-5} $. This constraint ensures that the density cavity solution remains physically stable and is used to define the parameter space explored in Figure \ref{fig:fpond}. For the path integral in Eq. (\ref{factor}), we integrate along the $z$-axis, accounting for multi-frequency contributions in the range $0.1\leq\omega_1\leq1$.

We evaluate $f_{\text{pond}}$ for a representative set of densities: $n_3 \in \{1,5,10,30\}$, with fixed temperature $T_4=0.1$ and perturbation amplitude $\alpha = 10^{-4}$.  Given that the scale of the SS is negligible compared to the macroscopic stellar wind variations, the background field $B_0$ is treated as locally homogeneous. Figure \ref{fig:fpond} (upper panel) illustrates the dependence of $f_{\text{pond}}$ on the transverse magnetic perturbation $B_{\theta 0}$. For comparison, the lower panel displays the quadratic correction factor $f^*_{\text{pond}}$, which governs the optical depth modification. For a single wave packet with ambient density $n_{e0} = 10^4 \, \text{cm}^{-3}$, background field $B_0 = 4.3\times 10^{-2}$ G, and a characteristic path length $L \approx 10^{10}~\text{cm}$, the baseline RM contribution is approximately ${\rm RM}_{\text{single}} \approx 1.15 \, \text{rad m}^{-2}$. This value serves as the reference unit for our cumulative RM estimates.

To address the sensitivity of our model to the wave amplitude and ambient conditions, we have expanded the parameter space exploration in Figure \ref{fig:fpond-contour}. This contour plot maps the ponderomotive filling factor $f_{\text{pond}}$ ($W_{0}$ branch) across a broad range of ambient magnetic fields $B_{0}$ and perturbation amplitudes $\alpha$, extending up to $\alpha \approx 0.3$ ($10^{-0.5}$). With a fixed electron density $n_{e0} = 10^{4}~\text{cm}^{-3}$, the results illustrate that significant density depletion is sustained across several orders of magnitude in $\alpha$. The unshaded regions denote regimes formally excluded by the requirement for real-valued density solutions, governed by the constraint $\alpha B_1 / (T_4^{1/2} n_3^{1/2}) \le 1.1 \times 10^{-5}$. This relation highlights that the permissible wave amplitude is intrinsically coupled to the background plasma temperature and density; in hotter or denser environments, the mechanism remains effective for much higher $\alpha$ values. This global perspective demonstrates that our fiducial model is not dependent on a fine-tuned value of $\alpha$, but represents a physically plausible slice within a wide and continuous parameter space where IAWs drive efficient plasma evacuation.

In the binary wind environment, IAWs are expected to populate the region where $r > r_{\text{IAW}}$. Triggered by beam instabilities or cascades, these waves can persist over a propagation distance determined by the Landau damping timescale, $D_{\text{prop}} \approx v_{\rm A} \tau_{\text{LD}} \sim 10^{12} \, \text{cm}$ (approximately $15 R_\odot$). For the typical parameters adopted here, we consider an annular region with inner radius $r_\text{IAW}$ and outer radius $v_\text{A}\tau_\text{LD}$ ( $r_\text{IAW} \ll v_\text{A}\tau_\text{LD}$), implying that the upper limit of the IAW propagation length is roughly 200 times the individual packet length $L$ ($D_{\text{prop}} \approx 200 L$). Consequently, the corresponding upper-limit contribution to the RM is 
\begin{eqnarray}
    \Delta\text{RM}\simeq-230 f_\text{pond}\left(\frac{B_0}{10^{-2}~\text{G}}\right)~ \text{rad}\cdot\text{m}^{-2},
    \label{result}
\end{eqnarray} 
evaluated at {$n_{e0}=10^{4}~\mathrm{cm}^{-3},~B_0=4.3\times10^{-2}~\text{G}$.}
As discussed above, we focus on the principal solution branch (the Lambert $W_{0}$ branch in Figure \ref{fig:fpond}), which corresponds to moderate density depletion. This choice is physically motivated by continuity: the system represents a perturbation of the background plasma, meaning the density must approach the ambient level ($n_e \to n_{e0}$) as the driving current vanishes ($\xi \to 0$). Adopting a representative filling factor of $f_{\text{pond}} \approx 0.17$ from this branch, we estimate a cumulative RM depletion of $\Delta {\rm RM} \approx -39 \, \text{rad m}^{-2}$. This magnitude is consistent with the observed ``dips". Physically, once the driving electron beam subsides, the IAW turbulence decays, and the RM is expected to recover to its ambient value.

To demonstrate the diagnostic utility of our model, we leverage the observed RM substructures to probe the plausible range of plasma conditions and magnetic activity levels associated with FRB sources. Figure \ref{fig:rm201124} displays the temporal RM evolution of FRB 20201124A \citep{2022Natur.609..685X} and FRB 20220529 \citep{doi:10.1126/science.adq3225}. To isolate short-term fluctuations, the secular trends in both sources are modeled using a composite fit consisting of a first-order polynomial and a sinusoidal component (representing orbital modulation), optimized via least-squares minimization.

For FRB 20201124A, the RM ranges from approximately $-300$ to $200 \, \text{rad m}^{-2}$, consistent with our fiducial background estimates ($n_{e0} \approx 10^4 \, \text{cm}^{-3}$, $B_0 \approx 4.3 \times 10^{-2}$ G). The regions highlighted by purple hatching serve as representative examples of the rapid RM suppression frequently observed in this source. Specifically, the first selected event shows a drop of $\Delta \rm RM_1 \approx -30 \, \text{rad m}^{-2}$ relative to the local baseline ($50\text{--}180 \, \text{rad m}^{-2}$), while the second event exhibits a deeper dip of $\Delta \rm RM_2 \approx -100 \, \text{rad m}^{-2}$. In both instances, the RM declines within $\sim 1$ day and recovers over $1\text{--}2$ days. Remarkably, the magnitude of the first event ($\approx -30 \, \text{rad m}^{-2}$) falls entirely within the predicted limit of our IAW-induced cavity model ($\sim -39 \, \text{rad m}^{-2}$), demonstrating that the fiducial model can naturally account for such variations without fine-tuning. For the second, deeper event, while the fiducial prediction is smaller, a moderate local enhancement in plasma parameters (e.g., magnetic field or clumpiness) can readily bridge the gap. We emphasize that these analyzed intervals are not unique; rather, they are illustrative of the ubiquitous short-term RM jitter present throughout the observational campaign. This widespread variability suggests that IAW-induced cavities may be a common and persistent feature of the local plasma environment.

In the case of FRB 20220529, the dynamic range of the RM ($-300$ to $300 \, \text{rad m}^{-2}$) is approximately 1.5 times broader than that of FRB 20201124A, implying a stronger ambient magnetic field or higher electron density. Prominent substructures are evident at multiple epochs, such as MJD 59830, 59870, and 59900. Notably, the extended region highlighted by the wide purple band encapsulates a cluster of successive day-scale dips, where the RM fluctuates by $\sim 60\text{--}200 \, \text{rad m}^{-2}$. Our estimated upper limit for the IAW-induced correction ($\sim -58 \, \text{rad m}^{-2}$) accounts for a significant fraction of this variability. The frequent occurrence and high amplitude of these excursions suggest that the local environment of FRB 20220529 is substantially more complex or extreme than that of FRB 20201124A, likely populated by multiple superimposed wave packets or regions of intense magnetic activity.

\section{Discussion and conclusions} \label{sec:discussion}

In this work, we have investigated the generation of plasma substructures via the ponderomotive force and evaluated their impact on the RM and DM of FRBs. While secular RM variations in repeating FRBs are often attributed to the orbital motion of binary systems, the superimposed short-term fluctuations require a localized, fast-acting mechanism. In the low-$\beta$ environment characteristic of massive stellar winds, large-scale \alfven~waves naturally cascade into IAWs through both linear and nonlinear process. Due to their finite parallel electric fields and transverse scales comparable to the electron skin depth, these waves exert a significant ponderomotive force, driving a nonlinear feedback loop that creates deep electron density cavities.

Analytically, we solved the fluid equations coupled with the quasi-neutrality condition, deriving a transcendental relation (solved via the Lambert W function) that links the density depletion to the wave current intensity. Applying this formalism to a binary wind scenario, we summarize our main findings:
\begin{itemize}
    \item Mechanism: The ponderomotive effect is highly efficient in expelling electrons from the wave core, creating density ``dips" that are bounded by minor ion density ridges. This provides a natural physical explanation for the rapid RM suppression events observed in active repeaters.
    \item Magnitude: For typical magnetar-massive star binary parameters  ($n_e \sim 10^4 \, \text{cm}^{-3}, B \sim 10^{-2}$ G), the cumulative effect of an extended IAW turbulence region ($\sim 15 R_\odot$) can induce RM variations of order $\Delta {\rm RM} \sim -40$ to $-60 \, \text{rad m}^{-2}$.
    \item Comparison with Observations: Comparing our model with data, we find that the specific ``dip" substructures ($\Delta {\rm RM} \approx -30 \, \text{rad m}^{-2}$) observed in FRB 20201124A are well-reproduced by our fiducial estimates. Conversely, FRB 20220529 exhibits extreme variability ($\left|\Delta {\rm RM}\right| > 100 \, \text{rad m}^{-2}$ in a single day) and a larger dynamic range. While our model accounts for a significant portion of these fluctuations, the extreme amplitude implies a local environment with either stronger magnetic fields, higher plasma densities, or more intense turbulence than the fiducial parameters assumed here.
\end{itemize}

We emphasize two distinct advantages of the model proposed here:

\begin{itemize}

\item Applicability: Short-term RM jitter is a common feature in active repeating FRBs. Our model provides a deterministic link between RM variations and kinetic plasma scales, which is viable across a broad regime of the ($n, T, B, \alpha$) parameter space. This framework remains a robust candidate for explaining such fluctuations under physically plausible conditions in magnetized binary winds.

\item Quantifiability: Unlike previous studies that rely on phenomenological or stochastic turbulence descriptions (e.g., random density fluctuations), we provide, for the first time, a quantitative estimation method based on first principles. This deterministic framework links the observed RM variations directly to the kinetic scales of the plasma, filling a gap in previous literature where such quantitative links were largely absent.
\end{itemize}

In conclusion, IAW-driven density cavitation offers a robust mechanism for generating the fine-scale RM substructures observed in FRB environments. Future high-cadence polarimetric monitoring of repeaters will be crucial to distinguishing this mechanism from other propagation effects (e.g., ecliptic passage or discrete clumps) and will serve as a unique probe of kinetic-scale turbulence in extragalactic environments.

\begin{acknowledgments}
Q.Z. thanks Xuan Yang and Songbo Zhang for kindly providing the observational data used in this work. This work is supported by the National Natural Science Foundation of China (Grant  Nos. 12373052, 12321003, 12393813) and the CAS Project for Young Scientists in Basic Research  (Grant No. YSBR-063).
\end{acknowledgments}

\bibliography{sample7}{}
\bibliographystyle{aasjournalv7}

\appendix

\section{Dispersive Relation for IAWs} \label{appendix:dispersion}

The dispersion relation for IAWs can be derived from the two-fluid momentum equations for species $j=\{e,i\}$
\begin{eqnarray}
    \p_t \b{u}_{j} + \left(\b{u}_{j} \cdot \nabla\right) \b{u}_{j} = \frac{q_j}{m_j}\left( \b{E} + \b{u}_{j} \times \b{B}\right) - \frac{1}{m_j n_{j}}\nabla P_j,
    \label{tfme:1}
\end{eqnarray}
combined with the continuity equation
\begin{eqnarray}
    \p_t n_{j} + \nabla \cdot \left(n_{j} \b{u}_{j}\right) = 0,
    \label{tfme:2}
\end{eqnarray}
and Maxwell's equations:
\begin{eqnarray}
    \nabla \times \b{E} &=& -\p_t \b{B},\\
    \label{tfme:3}
    \nabla \times \b{B} &=& \mu_0\b{J} + \frac{1}{c^2}\p_t \b{E}.
    \label{tfme:4}
\end{eqnarray}

We consider the low-$\beta$ limit where thermal pressure is negligible. We introduce the scalar potential $\phi$ and the vector potential $\b{A} = A_z \b{\hat{z}}$ in discussion of IAWs, where we have neglected the perpendicular component of vector potential. Thus, the magnetic field perturbation is $\b{B} = \nabla\times A_z \b{\hat{z}}$ and the electric field components are:
\begin{eqnarray}
    &&E_\parallel = -\p_z\phi - \p_t A_z,
    \label{elec:para}\\
    &&E_\perp = - \nabla_\perp \phi.
    \label{elec:perp}
\end{eqnarray}
The current density is related to the magnetic field via Eq.~(\ref{tfme:4}). Using vector identities $\nabla\times(\nabla\times A_z \b{\hat{z}}) = \nabla (\nabla \cdot A_z \b{\hat{z}}) - \nabla^2 A_z \b{\hat{z}}$, the parallel and perpendicular components of the current are:
\begin{eqnarray}
    &&\mu_0 J_\parallel = -\nabla_\perp^2 A_z,
    \label{para:curr}\\
    &&\mu_0 \b{J}_\perp = \nabla_\perp \frac{\p A_z}{\p z},
    \label{perp:curr}
\end{eqnarray}
Using $J_\parallel = n_e q_e u_{e\parallel}$ and Eq.~(\ref{para:curr}) we have
\begin{eqnarray}
    u_{e\parallel} = - \frac{1}{\mu_0 n_e q_e} \nabla_\perp^2 A_z,
    \label{u:para}
\end{eqnarray}
We now focus on the IAW regime. The electron dynamics along the field line is dominated by inertia, yielding the parallel momentum equation:
\begin{eqnarray}
     \p_t J_\parallel = n_e q_e \p_t u_{e\parallel} = \frac{n_e q_e^2}{m_e} E_\parallel,
    \label{current:para} 
\end{eqnarray}
In the perpendicular direction, the current is dominated by the ion polarization drift due to the large ion mass:
\begin{eqnarray}
     \b{J}_\perp = n_i q_i \b{u}_{i} = \frac{n_i m_i}{B^2} \frac{\d \b{E}_\perp}{\d t}.
    \label{current:perp}
\end{eqnarray}
Combine Eqs.~(\ref{elec:para})(\ref{u:para})(\ref{current:para}), we obtain the parallel wave equation:
\begin{eqnarray}
    &&\left(1-\frac{m_e}{\mu_0 n_e q_e^2}\nabla_\perp^2\right)\frac{\p A_z}{\p t} = -\frac{\p\phi}{\p z}.
    \label{vec:scal}
\end{eqnarray}
Similarly, combining Eqs.~(\ref{elec:perp})(\ref{perp:curr})(\ref{current:perp}) leads to
\begin{eqnarray}
    &&\frac{\p A_z}{\p z} = -\frac{\mu_0 n_i m_i}{B_0^2} \frac{\p\phi}{\p t}.
    \label{scal:vec}
\end{eqnarray}
Identifying the electron skin depth $\lambda_e^2 = m_e / (\mu_0 n_e e^2)$ and the \alfven~speed $v_{\rm A}^2 = B_0^2 / (\mu_0 n_i m_i)$, we combine Eqs.~(\ref{vec:scal}) and (\ref{scal:vec}) to derive the wave equation:
\begin{eqnarray}
    &&\left(1-\lambda_e^2\nabla_\perp^2\right) \frac{\p^2A_z}{\p t^2} = v_{\rm A}^2 \frac{\p^2 A_z}{\p z^2},
    \label{wave:IAW}
\end{eqnarray}
Assuming a plane wave solution $\exp[i(k_\parallel z + k_\perp x - \omega t)]$, we obtain the IAW dispersion relation:
\begin{eqnarray}
    &&\omega^2 = \frac{k_\parallel^2 v_{\rm A}^2}{1+k_\perp^2 \lambda_e^2}.
    \label{disp:IAW}
\end{eqnarray}
In the limit $k_\perp \lambda_e \ll 1$, this reduces to the standard shear \alfven~wave relation $\omega = k_\parallel v_{\rm A}$. The group velocity of IAW is given by: 
\begin{eqnarray}
    \mathbf{v}_g =\frac{\d \omega}{\d \b{k}} = \frac{v_{\rm A}}{\sqrt{1+k_\perp^2 \lambda_e^2}}\b{\hat{z}}  - \frac{\omega k_\perp\lambda_e^2 }{1+k_\perp^2 \lambda_e^2} \b{\hat{x}} .
    \label{group:IAW}
\end{eqnarray}

\end{document}